\documentclass[reprint,amsmath,amssymb,aps,prl,nobibnotes,nofootinbib]{revtex4-2}

\usepackage[utf8]{inputenc} 
\usepackage[T1]{fontenc}    
\usepackage{hyperref}       
\usepackage{url}            
\usepackage{booktabs}       
\usepackage{amsfonts}       
\usepackage{nicefrac}       
\usepackage{microtype}      
\usepackage{lipsum}
\usepackage{graphicx}
\usepackage{todonotes}

\usepackage[switch]{lineno}
 
\begin{document}

\preprint{preprint}

\title{Detecting long-lived particles trapped in detector material at the LHC}

\author{Jan Kieseler}
 \email{jan.kieseler@cern.ch}
\author{Juliette Alimena}
\author{Jasmine Simms}
 \altaffiliation[Also at ]{The University of Oxford\\ Oxford OX1 2JD, United Kingdom}
\author{Thea Aarrestad}
\author{Maurizio Pierini}
\affiliation{ European Center for Nuclear Research (CERN) \\ CH 1211, Geneva 23, Switzerland }

\author{Alexander Kish}
\affiliation{Fermi National Accelerator Laboratory\\
Batavia, IL 60510 USA}

\date{October 26, 2021}

\begin{widetext}
\begin{abstract}
We propose a two-stage strategy to search for new long-lived particles that could be produced at the CERN LHC, become trapped in detector material, and decay later. In the first stage, metal rods are exposed to LHC collisions in an experimental cavern. In the second stage, they are immersed in liquid argon at a different location, where out-of-time decays could be detected. Using a benchmark of pair-produced long-lived gluinos, we show that this experiment would have unique sensitivity to gluino-neutralino mass splittings down to 3~GeV, in previously uncovered lifetimes of days to years.
\end{abstract}
\end{widetext}

\maketitle


The search for new particles produced in proton-proton collisions is one of the main aspects of the physics program at the CERN LHC. One of the most intriguing scenarios explored in these searches for physics beyond the standard model is that of long-lived particles~\cite{Alimena:2019zri} produced in collisions and traveling in the detector before decaying. Depending on their lifetime and charge, these particles could generate a rich set of signatures in particle detectors: displaced vertices, track kinks, appearing jets, heavy muons, etc. These long-lived particles are being searched for with the ATLAS~\cite{Aad:2008zzm}, CMS~\cite{Chatrchyan:2008zzk}, and LHCb~\cite{LHCb:2008vvz} detectors at the LHC, and also with dedicated detectors such as FASER~\cite{FASER:2018eoc,FASER:2019aik}, milliQan~\cite{milliQan:2021lne}, and the planned CODEX-b~\cite{Aielli:2019ivi}. In one of the most extreme scenarios, heavy particles could be trapped in detector material and decay after some time, as originally proposed in Ref.~\cite{Asai:2009ka}. Typical examples of this kind include the production of long-lived sleptons in many SUSY scenarios, or long-lived gluinos ($\widetilde{g}$)~\cite{Arvanitaki:2005nq} predicted by Split SUSY~\cite{Wells:2003tf,Arkani_Hamed_2005,Giudice_2005}. As a benchmark example, we consider pair-production of the latter in this study. In this scenario, each long-lived gluino decays to a gluon (g) and a neutralino ($\widetilde{\chi}^0$), or to a quark-antiquark pair and a neutralino. Decays of these kinds were searched for by the CMS~\cite{Khachatryan:2015jha,Sirunyan:2017sbs} and ATLAS~\cite{ATL-PHYS-PUB-2019-019,Aad:2021yej} Collaborations, using triggers that fired in absence of colliding beams. Gluino masses ($m_{\widetilde{g}}$) lighter than 1.4~TeV were excluded, for gluino proper lifetimes ($\tau_{\widetilde{g}}$) of $10^{-5}$ to $10^3$~s, assuming the gluinos decay to a quark-antiquark pair and a neutralino with a mass ($m_{\widetilde{\chi}^0}$) of about 100~GeV, with 100\% branching fraction. Due to the nature of the trigger, these searches focused on large mass differences between the gluino and the neutralino. 

In this Letter, we propose to probe very long-lived scenarios, e.g., compressed gluino-neutralino spectra, with a dedicated detection strategy, consisting of two stages:
\begin{itemize}
    \item Trap: A removable inert material (RIM) is placed in the cavern of an LHC experiment. We assume that such a detector is CMS, which is the one with which we are most familiar. On the other hand, the strategy is general and could be adapted to other particle colliders and detectors at the collision points. For practical reasons, we consider brass rods, placed next to each other to form a block material with a shape optimized on specific aspects of the target scenarios, e.g., privileging $\eta$ coverage or absorption depth. Shielded by the particle detector, and by additional material if required, the RIM would receive only the most penetrating radiation, e.g. muons, neutrons, and potentially, new long-lived particles like gluinos. If the gluinos were moving sufficiently slowly, they could become trapped in the RIM.
    \item Detect: After the LHC run, the RIM is removed, the individual rods are separated, and they are placed in a cryostat filled with purified liquid argon (LAr). A voltage is applied to the rods, altering their polarity. We envision this basic LAr calorimeter setup, which could have simple readout electronics and is in line with current projects at CERN. Another potential detection setup could involve plastic scintillators and photosensors or similar calorimeter technologies. Fast muon timing detectors could be added to reject backgrounds from cosmic rays. When the long-lived gluinos decay, the energy of the decay products would be deposited in the LAr and the charge could be measured. We propose to use these technologies so that they could be shared or reused from ongoing CERN experiments.
\end{itemize}

The detection strategy is similar to the approaches discussed in Refs.~\cite{Feng:2004yi,DeRoeck:2005cur,Hamaguchi:2004df,Hamaguchi:2006vu}, where the trapping of charged sleptons in water tanks is described. The main difficulties with proposals involving water or LAr as the stopping material, which we considered in an early stage of this study, are that a large volume is required to reach an acceptable stopping rate, and the logistics are challenging (impossible) in the CMS (ATLAS) cavern. The strategy highlighted in this Letter provides several advantages: it exploits a more compact design; it comes with a movable target, which could cover different acceptance ranges during different exposure campaigns targeting different new physics scenarios; and it builds on many detector activities already ongoing at CERN, notably at the Neutrino Platform~\cite{Pietropaolo:2017jlh}. A similar trapping detector concept is also used by the MoEDAL Collaboration to look for trapped monopoloes and long-lived particles~\cite{MoEDAL:2020pyb,Acharya:2020uwc}.

Gluinos, which are the benchmark target of this study, can form strongly produced hadronic states called R-hadrons~\cite{Fairbairn:2006gg}. We simulate gluino R-hadrons that travel through a rough approximation of the CMS detector and approach a brass RIM with GEANT4~\cite{agostinelli2003geant4}. We use the Regge model to generate the R-hadron strong interactions with matter~\cite{Mackeprang:2006gx,Mackeprang:2009ad} and the FBERT physics list for the other processes. The CMS detector material is approximated with concentric cylinders. The innermost is made of air with a radius of 1~m to approximate the low-material silicon tracker, followed by led tungstate with a thickness of 20~cm to approximate the electromagnetic calorimeter, brass with a thickness of 1~m to approximate the hadronic calorimeter, and iron with a thickness of 2.9~m to approximate the iron in the CMS muon system. The RIM is modeled with 100 layers of brass, each of which are 2~cm thick in the $x$ direction and $2\times2$~m in the $y$-$z$ plane.

Neutral gluino R-hadrons are produced at the CMS beamspot and are shot directly at the brass RIM. Gluino masses from 5~GeV to 2.5~TeV are produced, and for each mass, we finely scan $\beta$, which is the gluino velocity divided by the speed of light, from 0.0 to 1.0. In total, we simulate 160 million gluino R-hadrons. 
We find that if the brass RIM was replaced by the same volume of LAr, the absorption efficiency is reduced by about a factor of 2, while if the same volume of water is used, the absorption efficiency is at least two orders of magnitude smaller.

Stable gluinos and gluino R-hadrons are generated with PYTHIA 8.306~\cite{PYTHIA8} in proton-proton collisions at a center-of-mass energy of 13~TeV, in a Split SUSY scenario. The gluinos are pair-produced through gluon-gluon fusion and quark-antiquark annihilation. We set the fraction of gluinos that hadronize into a gluino-gluon state, which is a free parameter in the hadronization model and determines the fraction of R-hadrons that are neutral at production, to be 0.1. We generate 100,000 events per gluino mass, which are the same masses as those that we simulate with GEANT4. 

The smallest gluino masses we consider have maximum angular acceptance near $\theta=0$ and $\pi$, while the largest gluino masses have the maximum acceptance near $\theta=\pi/2$. Therefore, we consider two RIM absorber positions, just outside of the CMS detector: the first positioned at $\eta=0$, covering $1.44<\theta<1.71$ ($P_0$); and the second at $\eta=2.3$, covering $0.18<\theta<0.35$ ($P_1$).

\begin{figure}[htb]
\includegraphics[width=0.45\textwidth]{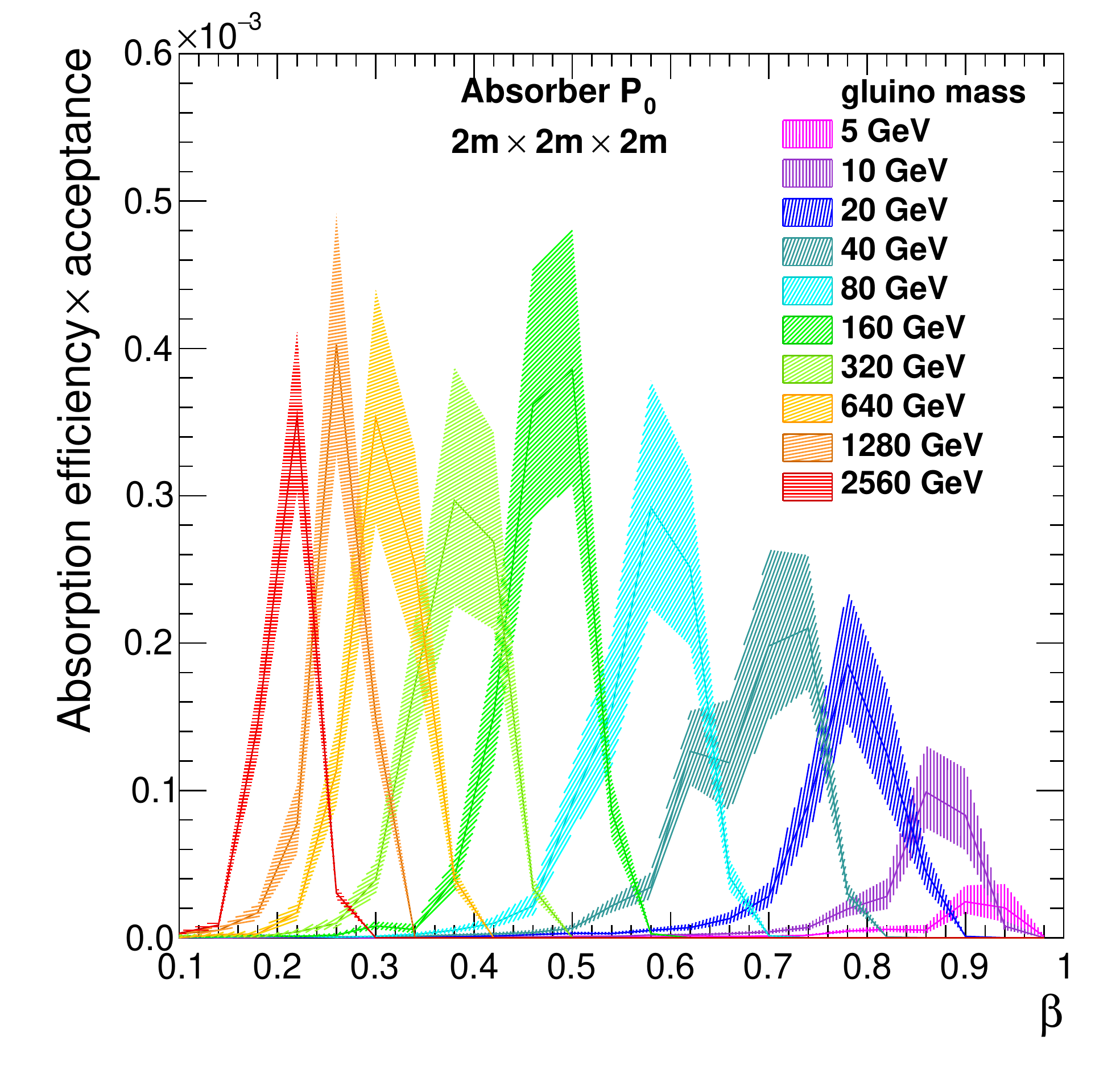}
\includegraphics[width=0.45\textwidth]{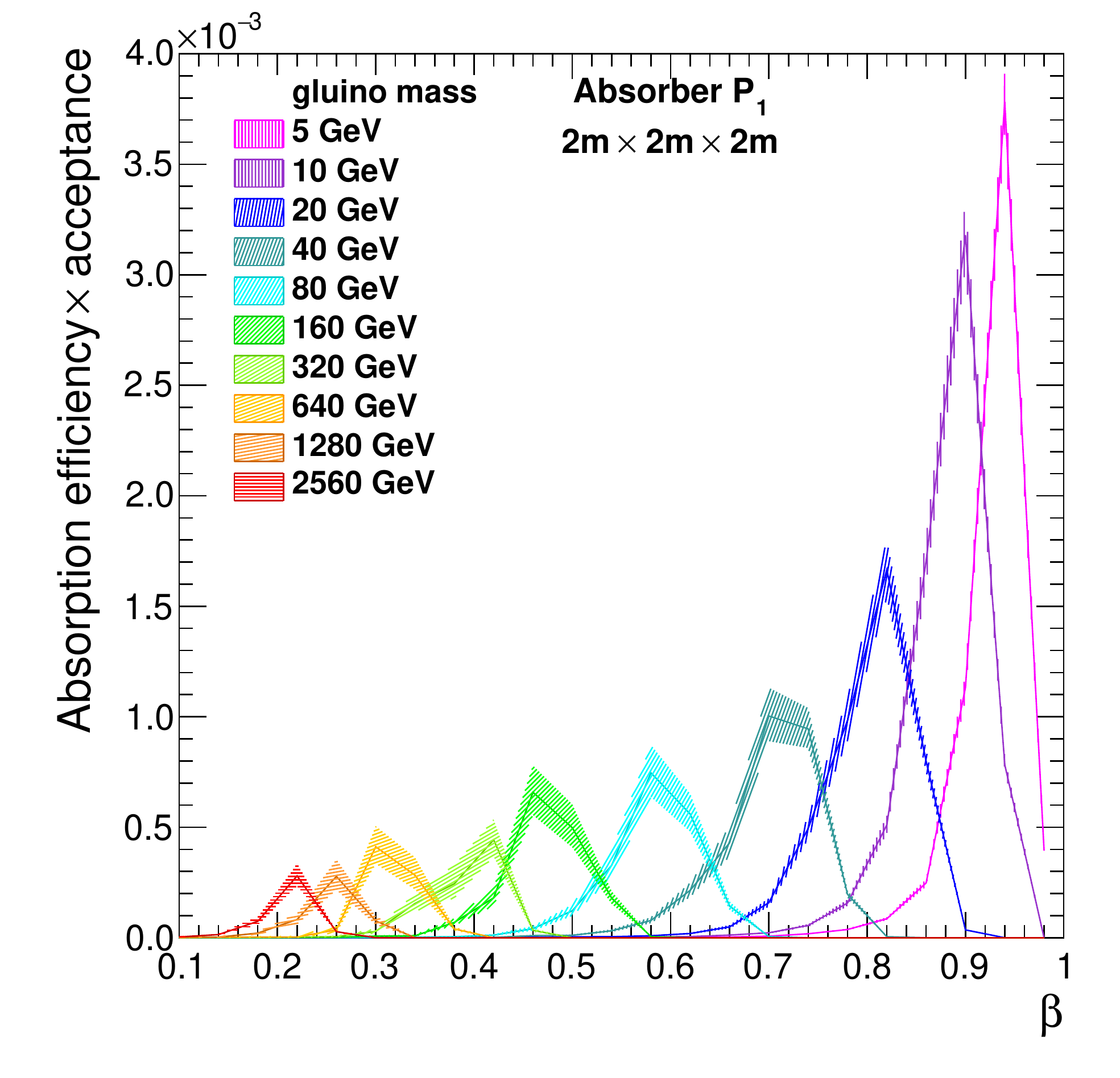}
\caption{Gluino R-hadron acceptance times the absorption efficiency as a function of $\beta$ for different gluino masses, for $P_0$ (top) and $P_1$ (bottom). An absorber size of 2~m$^3$ is assumed.
}
\label{fig:EffXAcc}
\end{figure}

We then convolve the R-hadron angular acceptance with the absorption efficiency in order to obtain the total efficiency times acceptance for the gluino R-hadrons to hit and be absorbed in the RIM, as shown as a function of $\beta$ in Fig.~\ref{fig:EffXAcc} for absorber $P_0$ (top) and $P_1$ (bottom). As a particle travels through CMS and the brass absorber, it will be slowed by material until it comes to a stop. For very small $\beta$ values, the particles are stopped before they reach the absorber, and for very large $\beta$ values the particles continue through the detector and the absorber without being stopped. Between these two extremes, there is a range of $\beta$ values for each $m_{\widetilde{g}}$ that will come to a stop within the absorber and therefore be absorbed by the brass. There are clear maxima at different $\beta$ values for the different $m_{\widetilde{g}}$ due to the absorption efficiency. For the absorber at $P_0$, the angular acceptance times efficiency is largest for $\beta\lesssim0.6$. For the absorber at $P_1$, the acceptance times efficiency for gluinos with masses greater than about 300~GeV is roughly the same as for $P_0$. However, there is a sharp increase in acceptance times efficiency for low mass gluinos, particularly for $m_{\widetilde{g}}\lesssim 50$~GeV. This is due to the low-mass gluinos having peak absorption efficiency at large $\beta$ values, which is where the angular acceptance is the highest for absorber at $P_1$. For example, for the absorber at $P_1$, gluinos with a mass of 5~GeV have a maximum acceptance times efficiency of 0.004 at $\beta=0.94$. The total acceptance times efficiency ranges from about $10^{-4}$ to $6 \cdot 10^{-3}$ for other $m_{\widetilde{g}}$.

We take a benchmark metal absorber size of $8~\text{m}^3$ and a benchmark rod spacing of 1~cm in LAr. Thus, the entire detection setup would be $32~\text{m}^3$ and the LAr volume would be $24~\text{m}^3$. We assume the detection setup is able to detect hadronic and electromagnetic activity with a total energy release of about 3~GeV and greater, as this is very well within the capabilities of typical LAr calorimeters. Further, we have verified in simulation that particles with momenta in the range of 100~MeV or higher can escape the rods and deposit energy in the sensitive material. The setup will be placed in an experimental hall with shielding from cosmic rays, but some muon showers from cosmic rays could still penetrate the setup. 

Cosmic ray muons with energies between 1~GeV and 50~TeV are simulated and propagated diagonally through the GEANT4 implementation of the brass rods immersed in LAr, which maximizes the material the muons traverse and provides a conservative assumption. This simulation shows that highly energetic muons can leave about 1\% of their energy in the LAr, for rod spacings of about 1~cm or more. We take the spectrum of vertical cosmic ray muons at sea level from Ref.~\cite{Bugaev:1998bi} as an upper limit on the total number of expected muons~\cite{Tsuji1998MeasurementsOM}, and convolve it with the fraction of energy that the cosmic ray muons leave in the detection setup. We determine a cosmic ray muon background estimate by integrating the convolved muon spectrum over the momentum, starting from a threshold. We assume this threshold at which the cosmic ray muons could mimic the signal jet in the absence of other rejection methods is $\frac{\Delta m}{2}$, where $\Delta m = m_{\widetilde{g}} - m_{\widetilde{\chi}^0}$, as half the energy of the gluino R-hadron decaying at rest will be detectable.

\begin{figure}[htb]
\includegraphics[width=0.5\textwidth]{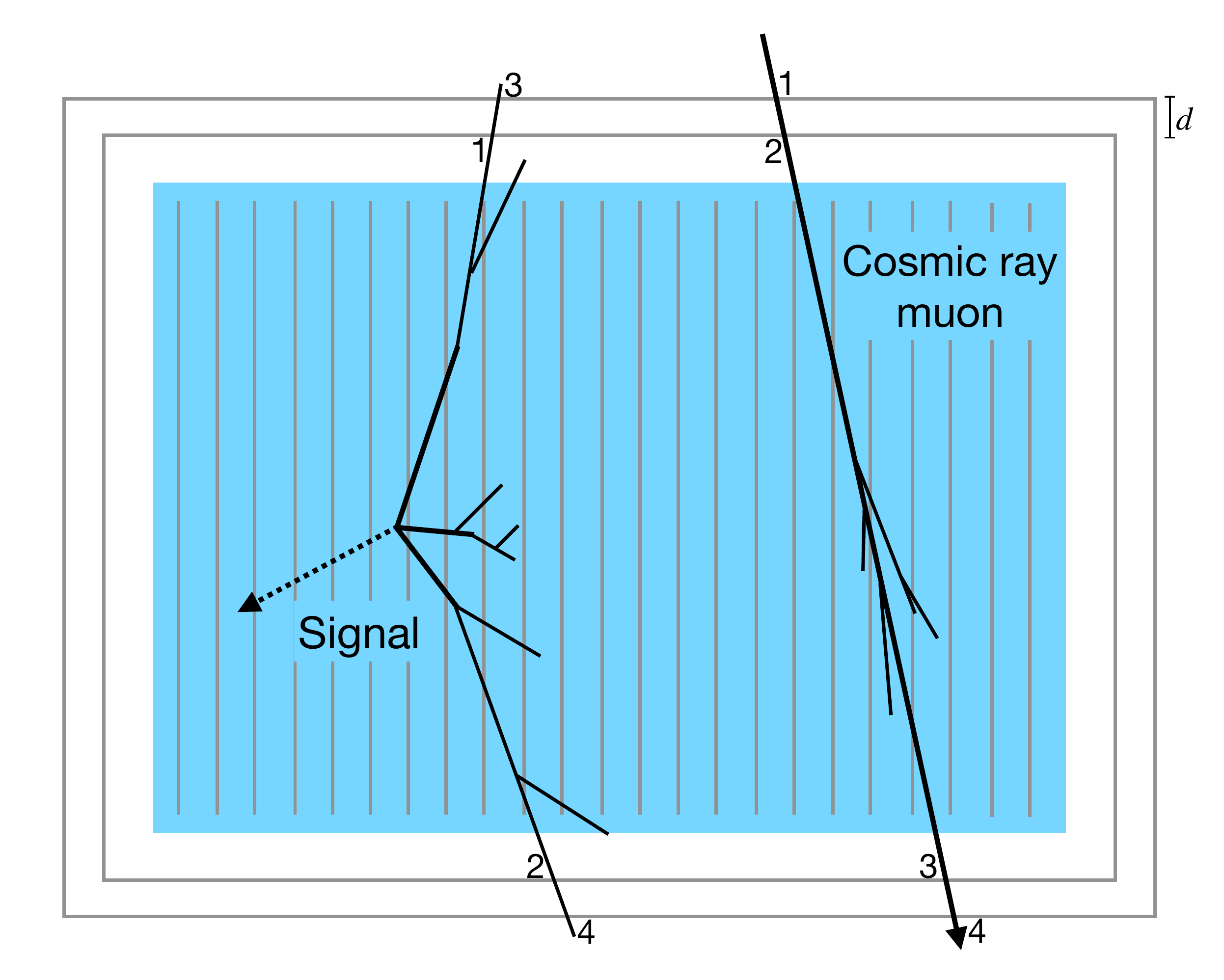}
\caption{The detection setup, with the brass rods shown with vertical gray lines, LAr shown in blue, and double layers of RPCs shown with the outermost gray lines. A sketch of a signal gluino decay is shown on the left, and a cosmic ray muon is shown on the right.
}
\label{fig:cosmicDetection}
\end{figure}

To reject the background from cosmic ray muons, we assume that a muon veto system with a fast response would be put in place. A high rejection power could be achieved using the timing capabilities of multilayer resistive plate chambers (RPCs) or plastic scintillators. For example, two layers of RPCs could be placed above and below the brass rods and LAr as shown in Fig.~\ref{fig:cosmicDetection}, spaced $d$~cm apart. A gluino decay is shown on the left side of the figure, and a muon from a cosmic shower is shown on the right. If the gluino decay produces a single shower, e.g., with a gluon in the final state, a signal would be easily distinguishable from a penetrating cosmic ray, since it would traverse the muon-veto system only once. If instead the gluino decay produces two showering quarks, particles from the two showers will interact with the innermost RPCs first and then the outermost RPCs, perhaps in the order as numbered in the figure. In contrast, a cosmic ray muon from the atmosphere will penetrate the setup from top to bottom, interacting with the RPCs in the order shown. Thus, if the RPCs have a timing resolution that is less than $d$ divided by the speed of light, muons from cosmic rays could be distinguished from showers from gluino decays.

There are several methods available to perform estimates of the cosmic muon background from data. First, the unexposed RIM could be inserted within the detection setup and the rate of cosmic muons could be measured for a period of time. In addition, the exposed RIM rods could be placed horizontally within the LAr instead of vertically and the detector readout could be performed in layers. This readout method would allow one to more carefully measure the direction of each shower and to determine if it was produced within the grid of rods or if it came from above. Furthermore, one could rely on a signal shape analysis to distinguish signal from background on a statistical basis. This could be, for example, a one-dimensional fit of signal and background templates or a multivariate classifier based on deep neural networks. We do not consider these possibilities here. In this respect, the background estimate presented in this paper should be considered as a first, conservative approximation.

\begin{figure}[htb]
\includegraphics[width=0.45\textwidth]{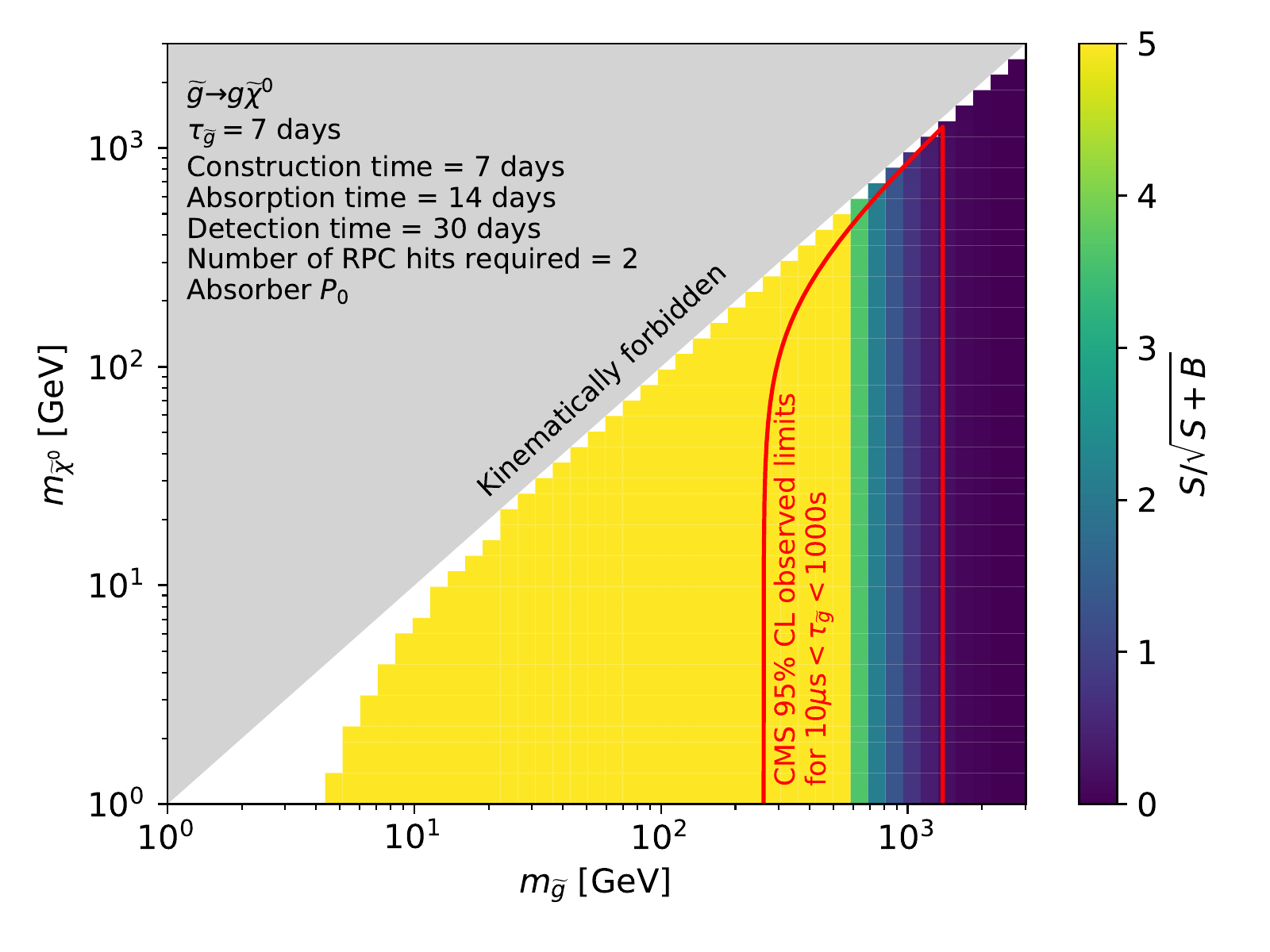}
\includegraphics[width=0.45\textwidth]{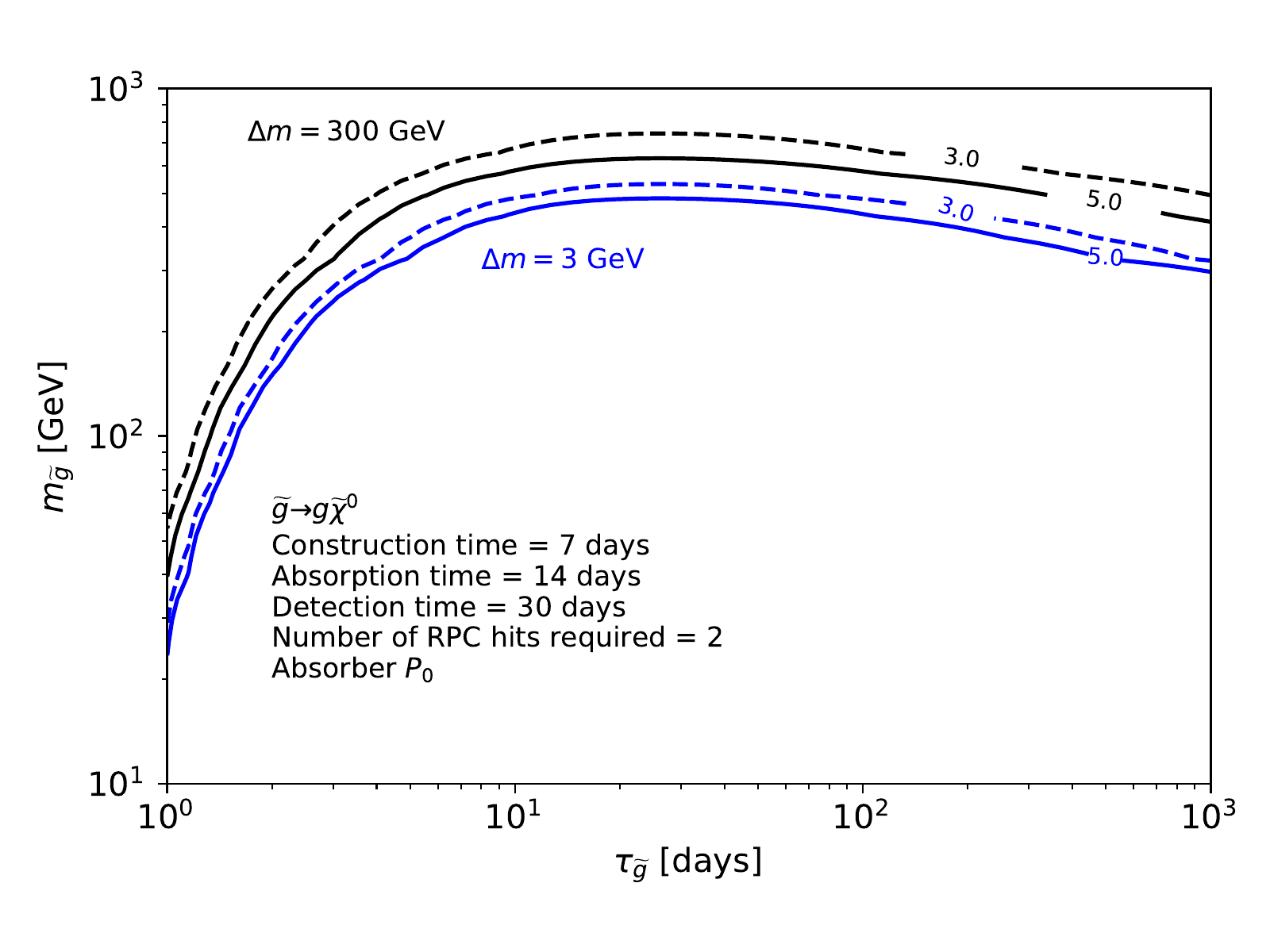}
\caption{The sensitivity of the experiment for a given neutralino mass and gluino mass is shown in the top plot. The grey area is kinematically forbidden. The 95\% confidence level observed upper limits from CMS on the gluino cross section times branching fraction for $10~\mu\text{s}<\tau_{\widetilde{g}}< 1000~\text{s}$~\cite{Sirunyan:2017sbs} are shown with a red line. The sensitivity of the experiment for a given gluino mass and proper lifetime is shown in the bottom plot, for $\Delta m = 3$~GeV (blue curves) and 300~GeV (black curves). The 3 and 5 standard deviation contours are shown with dashed and solid lines, respectively. In both plots, we assume a gluino proper lifetime of 7 days, 7 days to construct the detection setup, 14 days of absorption time, and 30 days of detection time. We also assume that 2 RPCs out of 4 detect the cosmic ray muons and that the absorber is in $P_0$.
}
\label{fig:sensitivity}
\end{figure}

The sensitivity of this experiment to the benchmark gluino signal, given the cosmic ray muon background estimate and expected number of gluino signal events described earlier, is shown in Fig.~\ref{fig:sensitivity}. The sensitivity is quantified as $S/\sqrt{S+B}$, where $S$ and $B$ are the number of signal and background events, respectively. We assume 99\% efficiency per muon detection layer, sufficient timing resolution with respect to the distance between each layer, and four layers in total. The minimum $\tau_{\widetilde{g}}$ this experiment would be sensitive to is defined by how long it will take to remove the RIM from the LHC experimental cavern and set it up in the LAr cryostat. For this process, we estimate about a week. The maximum $\tau_{\widetilde{g}}$ sensitivity is approximately how long we can run the detection for. As a result, the set of $\tau_{\widetilde{g}}$ that this experiment is sensitive to is complementary to that of ATLAS and CMS; these two experiments can probe $\tau_{\widetilde{g}}\lesssim 10$ days. Furthermore, since the calorimeter-like detection setup is sensitive to a few GeV in momentum, this experiment will be able to probe $\Delta m$ values of a few GeV. Thus, this experiment will have complementary mass coverage with respect to ATLAS and CMS as well.

We envision a possibility to use longer detection phases to probe more rare signals, such as particles charged only under the electromagnetic force. There could be one exposure campaign per year, of progressively longer length, and an appropriate data-acquisition period.

We assume repurposing one of the existing facilities at the CERN Neutrino Platform to make use of existing cryostat, purification, and electronic systems. This would substantially reduce the overall cost of this experiment, which is mainly driven by the LAr, the brass for the target, and the muon-veto system. We estimate the cost of the LAr would be ${\mathcal O}(10\mathrm{K})$ CHF (24K CHF, assuming that 1 kg of LAr corresponds to 0.7 liters and costs 0.7 CHF). We again assume $1~\text{cm}\times1~\text{cm}\times2~\text{m}$ rods and 200 rods in one dimension. Furthermore, we expect the price of the brass to be ${\mathcal O}(100 \mathrm{K})$ CHF (assuming that the price of brass is 4 CHF/kg, the total cost of the brass rods for the configuration described above would be about 280K CHF). This experiment considers cm-long drifts for ionization charge from particles with ${\mathcal O}(1)$ GeV energy or more, hence commercial grade argon with O$_{2}$, H$_{2}$O, or N$_{2}$ impurities at the ppb level will be acceptable, and no specific filtering or circulation system is necessary. We estimate that the cost of streamer-mode RPCs will have a small impact on the total, even though a precise quantification is not easy at this stage. It might also be possible to further reduce this cost by recycling old RPCs or spares from other experiments. In conclusion, we estimate that this experiment would cost about $\mathcal{O}(1\mathrm{M})$ CHF, including a factor of 2 safety margin to cover adapting the cryostat, building the electronics, and other contingencies.

In summary, we have proposed a two-stage experiment to discover new long-lived particles that could be produced at the LHC, stop in detector material, and decay later. Compared to a typical high-energy physics experiment, this experiment has the advantage of a relatively low cost and the possibility of a discovery reach within a few months of operation. The construction could be carried out without interfering with the existing scientific operations at CERN. See Supplemental Material at ~\cite{suppMaterial} for additional details and figures. This experiment would bring unique sensitivity to the small mass splitting regime, that is, from about 3 to 100~GeV. It would also be uniquely sensitive to lifetimes on the order of days to years.

\section*{Acknowledgements}
The seed of this study was a conversation witnessed by M.P. between Giacomo Polesello and Mihoko Nojiri at a coffee break of SUSY06 about the issues with fitting the water detectors proposed in Refs.~\cite{Feng:2004yi,DeRoeck:2005cur} in the LHC experiment halls. 

We thank Filip Moortgat for useful conversations at an early stage of this work, notably about Refs.~\cite{Feng:2004yi,DeRoeck:2005cur}.

A.K. is supported by Fermi Research Alliance, LLC under Contract No. DE-AC02-07CH11359 with the U.S. Department of Energy, Office of Science, Office of High Energy Physics.

\bibliographystyle{bibstyle}  
\bibliography{references}

\begin{thebibliography}{10}
\newcommand{\enquote}[1]{``#1''}
\providecommand{\url}[1]{\texttt{#1}}
\providecommand{\urlprefix}{URL }
\expandafter\ifx\csname urlstyle\endcsname\relax
  \providecommand{\doi}[1]{doi:\discretionary{}{}{}#1}\else
  \providecommand{\doi}{doi:\discretionary{}{}{}\begingroup
  \urlstyle{rm}\Url}\fi
\providecommand{\bibinfo}[2]{#2}
\providecommand{\eprint}[2][]{\url{#2}}

\bibitem{Alimena:2019zri}
\bibinfo{author}{J.~Alimena}, et~al., \enquote{\bibinfo{title}{{Searching for
  long-lived particles beyond the standard model at the Large Hadron
  Collider}},} \emph{\bibinfo{journal}{J. Phys. G}}, \bibinfo{volume}{47}
  \bibinfo{pages}{090501} (\bibinfo{year}{2020}),
  \doi{10.1088/1361-6471/ab4574}, \eprint{1903.04497}.

\bibitem{Aad:2008zzm}
\bibinfo{author}{{ATLAS Collaboration}}, \enquote{\bibinfo{title}{{The ATLAS
  experiment at the CERN Large Hadron Collider}},}
  \emph{\bibinfo{journal}{JINST}}, \bibinfo{volume}{3} \bibinfo{pages}{S08003}
  (\bibinfo{year}{2008}), \doi{10.1088/1748-0221/3/08/S08003}.

\bibitem{Chatrchyan:2008zzk}
\bibinfo{author}{{CMS Collaboration}}, \enquote{\bibinfo{title}{The {CMS}
  experiment at the {CERN} {LHC}},} \emph{\bibinfo{journal}{JINST}},
  \bibinfo{volume}{3} \bibinfo{pages}{S08004} (\bibinfo{year}{2008}),
  \doi{10.1088/1748-0221/3/08/S08004}.

\bibitem{LHCb:2008vvz}
\bibinfo{author}{{LHCb Collaboration}}, \enquote{\bibinfo{title}{{The LHCb
  detector at the LHC}},} \emph{\bibinfo{journal}{JINST}}, \bibinfo{volume}{3}
  \bibinfo{pages}{S08005} (\bibinfo{year}{2008}),
  \doi{10.1088/1748-0221/3/08/S08005}.

\bibitem{FASER:2018eoc}
\bibinfo{author}{{FASER Collaboration}},
  \enquote{\bibinfo{title}{{FASER\textquoteright{}s physics reach for
  long-lived particles}},} \emph{\bibinfo{journal}{Phys. Rev. D}},
  \bibinfo{volume}{99} \bibinfo{pages}{095011} (\bibinfo{year}{2019}),
  \doi{10.1103/PhysRevD.99.095011}, \eprint{1811.12522}.

\bibitem{FASER:2019aik}
\bibinfo{author}{{FASER Collaboration}}, \enquote{\bibinfo{title}{{FASER:
  ForwArd Search ExpeRiment at the LHC}},}  (\bibinfo{year}{2019}),
  \eprint{1901.04468}.

\bibitem{milliQan:2021lne}
\bibinfo{author}{{milliQan Collaboration}},
  \enquote{\bibinfo{title}{{Sensitivity to millicharged particles in future
  proton-proton collisions at the LHC with the milliQan detector}},}
  \emph{\bibinfo{journal}{Phys. Rev. D}}, \bibinfo{volume}{104}
  \bibinfo{pages}{032002} (\bibinfo{year}{2021}),
  \doi{10.1103/PhysRevD.104.032002}, \eprint{2104.07151}.

\bibitem{Aielli:2019ivi}
\bibinfo{author}{G.~Aielli}, et~al., \enquote{\bibinfo{title}{{Expression of
  interest for the CODEX-b detector}},} \emph{\bibinfo{journal}{Eur. Phys. J.
  C}}, \bibinfo{volume}{80} \bibinfo{pages}{1177} (\bibinfo{year}{2020}),
  \doi{10.1140/epjc/s10052-020-08711-3}, \eprint{1911.00481}.

\bibitem{Asai:2009ka}
\bibinfo{author}{S.~Asai}, \bibinfo{author}{K.~Hamaguchi},
  \bibinfo{author}{S.~Shirai}, \enquote{\bibinfo{title}{{Measuring lifetimes of
  long-lived charged massive particles stopped in LHC detectors}},}
  \emph{\bibinfo{journal}{Phys. Rev. Lett.}}, \bibinfo{volume}{103}
  \bibinfo{pages}{141803} (\bibinfo{year}{2009}),
  \doi{10.1103/PhysRevLett.103.141803}, \eprint{0902.3754}.

\bibitem{Arvanitaki:2005nq}
\bibinfo{author}{A.~Arvanitaki}, et~al., \enquote{\bibinfo{title}{{Stopping
  gluinos}},} \emph{\bibinfo{journal}{Phys. Rev. D}}, \bibinfo{volume}{76}
  \bibinfo{pages}{055007} (\bibinfo{year}{2007}),
  \doi{10.1103/PhysRevD.76.055007}, \eprint{hep-ph/0506242}.

\bibitem{Wells:2003tf}
\bibinfo{author}{J.~D. Wells}, \enquote{\bibinfo{title}{{Implications of
  supersymmetry breaking with a little hierarchy between gauginos and
  scalars}},} \enquote{\bibinfo{booktitle}{{11th International Conference on
  Supersymmetry and the Unification of Fundamental Interactions}},}
  (\bibinfo{year}{2003}), \eprint{hep-ph/0306127}.

\bibitem{Arkani_Hamed_2005}
\bibinfo{author}{N.~Arkani-Hamed}, \bibinfo{author}{S.~Dimopoulos},
  \enquote{\bibinfo{title}{Supersymmetric unification without low energy
  supersymmetry and signatures for fine-tuning at the {LHC}},}
  \emph{\bibinfo{journal}{Journal of High Energy Physics}},
  \bibinfo{volume}{2005} \bibinfo{pages}{073–073} (\bibinfo{year}{2005}),
  ISSN \bibinfo{issn}{1029-8479}, \doi{10.1088/1126-6708/2005/06/073},
  \urlprefix\url{http://dx.doi.org/10.1088/1126-6708/2005/06/073}.

\bibitem{Giudice_2005}
\bibinfo{author}{G.~Giudice}, \bibinfo{author}{A.~Romanino},
  \enquote{\bibinfo{title}{Erratum to: “Split supersymmetry” [Nucl. Phys. B
  699 (2004) 65]},} \emph{\bibinfo{journal}{Nuclear Physics B}},
  \bibinfo{volume}{706} \bibinfo{pages}{487} (\bibinfo{year}{2005}), ISSN
  \bibinfo{issn}{0550-3213}, \doi{10.1016/j.nuclphysb.2004.11.048},
  \urlprefix\url{http://dx.doi.org/10.1016/j.nuclphysb.2004.11.048}.

\bibitem{Khachatryan:2015jha}
\bibinfo{author}{{CMS Collaboration}}, \enquote{\bibinfo{title}{{Search for
  decays of stopped long-lived particles produced in proton\textendash{}proton
  collisions at $\sqrt{s}= 8\,\text {TeV} $}},} \emph{\bibinfo{journal}{Eur.
  Phys. J. C}}, \bibinfo{volume}{75} \bibinfo{pages}{151}
  (\bibinfo{year}{2015}), \doi{10.1140/epjc/s10052-015-3367-z},
  \eprint{1501.05603}.

\bibitem{Sirunyan:2017sbs}
\bibinfo{author}{{CMS Collaboration}}, \enquote{\bibinfo{title}{{Search for
  decays of stopped exotic long-lived particles produced in proton-proton
  collisions at $\sqrt{s}=$ 13 TeV}},} \emph{\bibinfo{journal}{JHEP}},
  \bibinfo{volume}{05} \bibinfo{pages}{127} (\bibinfo{year}{2018}),
  \doi{10.1007/JHEP05(2018)127}, \eprint{1801.00359}.

\bibitem{ATL-PHYS-PUB-2019-019}
\bibinfo{author}{{ATLAS Collaboration}}, \enquote{\bibinfo{title}{{Generation
  and simulation of $R$-hadrons in the ATLAS experiment}},}
  \bibinfo{type}{Technical Report} \bibinfo{number}{ATL-PHYS-PUB-2019-019},
  \bibinfo{institution}{CERN}, \bibinfo{address}{Geneva}
  (\bibinfo{year}{2019}), \urlprefix\url{https://cds.cern.ch/record/2676309}.

\bibitem{Aad:2021yej}
\bibinfo{author}{{ATLAS Collaboration}}, \enquote{\bibinfo{title}{{A search for
  the decays of stopped long-lived particles at $\sqrt{s}=13$ TeV with the
  ATLAS detector}},} \emph{\bibinfo{journal}{JHEP}}, \bibinfo{volume}{07}
  \bibinfo{pages}{173} (\bibinfo{year}{2021}), \doi{10.1007/JHEP07(2021)173},
  \eprint{2104.03050}.

\bibitem{Feng:2004yi}
\bibinfo{author}{J.~L. Feng}, \bibinfo{author}{B.~T. Smith},
  \enquote{\bibinfo{title}{{Slepton trapping at the large hadron and
  international linear colliders}},} \emph{\bibinfo{journal}{Phys. Rev. D}},
  \bibinfo{volume}{71} \bibinfo{pages}{015004} (\bibinfo{year}{2005}),
  \doi{10.1103/PhysRevD.71.015004}, \bibinfo{note}{[Erratum: Phys.Rev.D 71,
  019904 (2005)]}, \eprint{hep-ph/0409278}.

\bibitem{DeRoeck:2005cur}
\bibinfo{author}{A.~De~Roeck}, et~al., \enquote{\bibinfo{title}{{Supersymmetric
  benchmarks with non-universal scalar masses or gravitino dark matter}},}
  \emph{\bibinfo{journal}{Eur. Phys. J. C}}, \bibinfo{volume}{49}
  \bibinfo{pages}{1041} (\bibinfo{year}{2007}),
  \doi{10.1140/epjc/s10052-006-0182-6}, \eprint{hep-ph/0508198}.

\bibitem{Hamaguchi:2004df}
\bibinfo{author}{K.~Hamaguchi}, \bibinfo{author}{Y.~Kuno},
  \bibinfo{author}{T.~Nakaya}, \bibinfo{author}{M.~M. Nojiri},
  \enquote{\bibinfo{title}{{A Study of late decaying charged particles at
  future colliders}},} \emph{\bibinfo{journal}{Phys. Rev. D}},
  \bibinfo{volume}{70} \bibinfo{pages}{115007} (\bibinfo{year}{2004}),
  \doi{10.1103/PhysRevD.70.115007}, \eprint{hep-ph/0409248}.

\bibitem{Hamaguchi:2006vu}
\bibinfo{author}{K.~Hamaguchi}, \bibinfo{author}{M.~M. Nojiri},
  \bibinfo{author}{A.~de~Roeck}, \enquote{\bibinfo{title}{{Prospects to study a
  long-lived charged next lightest supersymmetric particle at the LHC}},}
  \emph{\bibinfo{journal}{JHEP}}, \bibinfo{volume}{03} \bibinfo{pages}{046}
  (\bibinfo{year}{2007}), \doi{10.1088/1126-6708/2007/03/046},
  \eprint{hep-ph/0612060}.

\bibitem{Pietropaolo:2017jlh}
\bibinfo{author}{F.~Pietropaolo}, \enquote{\bibinfo{title}{{Review of
  liquid-argon detectors development at the CERN neutrino platform}},}
  \emph{\bibinfo{journal}{J. Phys. Conf. Ser.}}, \bibinfo{volume}{888}
  \bibinfo{pages}{012038} (\bibinfo{year}{2017}),
  \doi{10.1088/1742-6596/888/1/012038}.

\bibitem{MoEDAL:2020pyb}
\bibinfo{author}{{MoEDAL Collaboration}}, \enquote{\bibinfo{title}{{First
  search for dyons with the full MoEDAL trapping detector in 13 TeV $pp$
  collisions}},} \emph{\bibinfo{journal}{Phys. Rev. Lett.}},
  \bibinfo{volume}{126} \bibinfo{pages}{071801} (\bibinfo{year}{2021}),
  \doi{10.1103/PhysRevLett.126.071801}, \eprint{2002.00861}.

\bibitem{Acharya:2020uwc}
\bibinfo{author}{B.~S. Acharya}, et~al., \enquote{\bibinfo{title}{{Prospects of
  searches for long-lived charged particles with MoEDAL}},}
  \emph{\bibinfo{journal}{Eur. Phys. J. C}}, \bibinfo{volume}{80}
  \bibinfo{pages}{572} (\bibinfo{year}{2020}),
  \doi{10.1140/epjc/s10052-020-8093-5}, \eprint{2004.11305}.

\bibitem{Fairbairn:2006gg}
\bibinfo{author}{M.~Fairbairn}, et~al., \enquote{\bibinfo{title}{{Stable
  massive particles at colliders}},} \emph{\bibinfo{journal}{Phys. Rept.}},
  \bibinfo{volume}{438} \bibinfo{pages}{1} (\bibinfo{year}{2007}),
  \doi{10.1016/j.physrep.2006.10.002}, \eprint{hep-ph/0611040}.

\bibitem{agostinelli2003geant4}
\bibinfo{author}{S.~Agostinelli}, et~al., \enquote{\bibinfo{title}{{GEANT4: a
  simulation toolkit}},} \emph{\bibinfo{journal}{Nucl. Instrum. Meth. A}},
  \bibinfo{volume}{506} (\bibinfo{year}{2003}),
  \doi{10.1016/S0168-9002(03)01368-8}.

\bibitem{Mackeprang:2006gx}
\bibinfo{author}{R.~Mackeprang}, \bibinfo{author}{A.~Rizzi},
  \enquote{\bibinfo{title}{{Interactions of coloured heavy stable particles in
  matter}},} \emph{\bibinfo{journal}{Eur. Phys. J. C}}, \bibinfo{volume}{50}
  \bibinfo{pages}{353} (\bibinfo{year}{2007}),
  \doi{10.1140/epjc/s10052-007-0252-4}, \eprint{hep-ph/0612161}.

\bibitem{Mackeprang:2009ad}
\bibinfo{author}{R.~Mackeprang}, \bibinfo{author}{D.~Milstead},
  \enquote{\bibinfo{title}{{An updated description of heavy-hadron interactions
  in GEANT4}},} \emph{\bibinfo{journal}{Eur. Phys. J. C}}, \bibinfo{volume}{66}
  \bibinfo{pages}{493} (\bibinfo{year}{2010}),
  \doi{10.1140/epjc/s10052-010-1262-1}, \eprint{0908.1868}.

\bibitem{PYTHIA8}
\bibinfo{author}{T.~Sj{\"o}strand}, et~al., \enquote{\bibinfo{title}{An
  introduction to {PYTHIA} 8.2},} \emph{\bibinfo{journal}{Comput. Phys.
  Commun.}}, \bibinfo{volume}{191} \bibinfo{pages}{159} (\bibinfo{year}{2015}),
  \doi{10.1016/j.cpc.2015.01.024}, \eprint{1410.3012}.

\bibitem{Bugaev:1998bi}
\bibinfo{author}{E.~V. Bugaev}, et~al., \enquote{\bibinfo{title}{{Atmospheric
  muon flux at sea level, underground and underwater}},}
  \emph{\bibinfo{journal}{Phys. Rev. D}}, \bibinfo{volume}{58}
  \bibinfo{pages}{054001} (\bibinfo{year}{1998}),
  \doi{10.1103/PhysRevD.58.054001}, \eprint{hep-ph/9803488}.

\bibitem{Tsuji1998MeasurementsOM}
\bibinfo{author}{S.~Tsuji}, et~al., \enquote{\bibinfo{title}{{Measurements of
  muons at sea level}},} \emph{\bibinfo{journal}{J. Phys. G}},
  \bibinfo{volume}{24} \bibinfo{pages}{1805} (\bibinfo{year}{1998}),
  \doi{10.1088/0954-3899/24/9/013}.

\bibitem{suppMaterial}
\enquote{\bibinfo{title}{Supplemental material for this paper},}
  (\bibinfo{year}{2021}).

\end{thebibliography}


\end{document}